\documentclass{appolb}
\usepackage{graphicx}
\begin{document}
\title{The feasibility study of the long-baseline neutrino oscillation experiment at the SUNLAB laboratory in Poland%
%%\thanks{ Presented at ...}%
}
\author{Ma\l{}gorzata Hara\'{n}czyk
\address{ Institute of Nuclear Physics Polish Academy of Sciences,\\  ul.Radzikowskiego 152, Krak\'{o}w} }
\maketitle
\begin{abstract}
The feasibility study of an underground laboratory in the Polkowice - Sieroszowice mine in Poland (SUNLAB) as a host of a far detector in a long-baseline neutrino oscillation experiment was performed. The SUNLAB location was previously studied under the LAGUNA FP7 project as a location for the underground multipurpose laboratory. The complementary study of the long-baseline neutrino experiment presented in this paper was performed as a continuation of this idea.  A neutrino beam produced at CERN and a far LAr-TPC detector hosted in the SUNLAB laboratory were simulated. The sensitivity of such an experiment for the determination of the CP symmetry violation in neutrino sector was calculated.
 The experiment at SUNLAB equipped with the 100 kton LAr TPC detector after 10 years of data taking can give the coverage of $\delta_{CP}$ parameter space of 58\% (60\%) for the normal (inverted) neutrino mass hierarchy at $3\sigma$ level and in both cases reaches $5\sigma$ level in case of the maximal violation. 

\end{abstract}
\section{Introduction}

The main purpose of the work presented in this paper was to explore the opportunities which would
be provided by the Sieroszowice Underground Laboratory SUNLAB, located in the
Polkowice-Sieroszowice mine in Poland, to study the CP symmetry conservation or violation in the neutrino sector.  The discovery of CP violation would be a milestone on the road to understand the observed dominance of matter over antimatter in the Universe, which is one of the fundamental questions of contemporary science. The discovery of neutrino oscillations in the years 1998-2002 and their interpretation within the framework of three neutrino flavour mixing, awarded the 2015 Nobel Prize in physics, provides the firm evidence for new physics beyond the Standard Model. In 2012, the $\theta_{13}$ mixing angle has been determined to be moderately large \cite{Daya_Bay},\cite{RENO},\cite{T2K_theta13}. This made it possible to use conventional high power neutrino beams in accelerator long baseline experiments to answer the remaining questions in neutrino oscillations, which are the neutrino mass hierarchy and the measurement of the CP-violating phase $\delta_{CP}$.

At present, the most important accelerator-based long baseline oscillation experiments are the T2K experiment in Japan \cite{T2K} and the NOvA experiment in the USA\cite{Nova}. If the CP violation is close to maximal, these two experiments should be able to determine $\delta_{CP}$ with a significance of three standard deviations within a decade. 
Concerning the future experiments, currently there are only two projects in the
world with neutrino beams of much higher intensities and much larger, more precise detectors, which are eligible: the LBNF-DUNE project at Fermilab in the USA\cite{LBNF_DUNE}, based on the liquid Argon detector technology, and the HyperKamiokande \cite{HyperK} giant water Cherenkov detector project, coupled with a beam from JPARC in Japan.

Feasibility studies of the future large underground laboratory in Europe, hosting huge detectors with a vast research programme including the accelerator long baseline oscillation studies, were performed within the EU FP7 LAGUNA project (2008-2011)\cite{LAGUNA}, followed by the EU FP7 LAGUNA-LBNO project (2011-2013)\cite{LAGUNA_LBNO}. SUNLAB was one of seven locations considered by LAGUNA. Long baseline oscillation studies of the CP violation discovery potential, with a neutrino beam from CERN and a huge liquid Argon time projection chamber (LAr-TPC) located at SUNLAB have been performed to conclude the feasibility study of the SUNLAB laboratory. The full description of the simulations of the long-baseline neutrino oscillation experiment at SUNLAB can be found in the PhD thesis \cite{PhD_MH}.

%%%%%%%%%%%%%%%%%%%%%%%%%%%%%%%%%%%%%%%%%%
\section{Studies for the SUNLAB location }
\label{sec:location}

The Sieroszowice UNderground LABoratory (SUNLAB) is a Polish R\&D project for building an underground laboratory in the Polkowice-Sieroszowice mine belonging to the KGHM holding of copper mines and metallurgic plants. The mine is located in south-western Poland in a distance of 950 km from CERN. The idea of such a laboratory was related to the initiative to look for a possible European location for a huge liquid Argon detector for studies of neutrino properties, neutrino astrophysics and searches for proton decays, developed within the European FP7 LAGUNA project. When LAGUNA started in 2008, the Polish site in the Polkowice - Sieroszowice mine, named SUNLAB, was one of the seven locations considered as future hosts for this large underground laboratory. Geomechanical feasibility studies for the SUNALB laboratory, hosting a 100 kton liquid Argon GLACIER detector in anhydrite rock, were performed in the years 2008-2011 within LAGUNA and gave very positive results \cite{LAGUNA_AZ}, \cite{LAGUNA_Pytel}. The best location to host a huge detector cavern resulted from the study was in a thick, stable layer of anhydrite at a depth of 665 m.

One should add that apart from anhydrite layers there are thick deposits of salt rock present in the Polkowice-Sieroszowice mine, characterised by very low level of natural radioactivity. This was shown  by the in situ gamma measurements and the radiochemical analysis of the rock samples demonstrating extremely low levels of radioactive isotopes of Uranium, Thorium and Potassium. The detailed description of results of natural radioactivity measurements in Sieroszowice mine can be found in  \cite{gamma_JKisiel}. However, as shown by the geomechanical simulations, the salt rock is too soft for excavation of huge caverns needed for the giant 100 kton detector. Nevertheless, an experimental activity using existing salt caverns on the 950 m.b.s. level with the extremely low background environment have been already performed \cite{HPGe_KPolaczek}.  The plan to build a low background underground laboratory in Polkowice -Sieroszowice salt rock is still under consideration.

Although the small laboratory in salt rock is an exciting project, the neutrino long baseline experiment coupled with the neutrino beam from CERN was the main motivation for the European FP7 LAGUNA project. Therefore the dedicated study for SUNLAB as a host of the far detector for the long baseline experiment with a baseline of 950 km was needed and performed within a grant of the Polish National Science Centre.

\vspace{2ex}%
%%%%%%%%%%%%%%%%%%%%%%%%%%%%%%%%%%%%%%%%%%%%%

\section{ Neutrino oscillations}
\label{sec:neutrino}

The approximate analytic expression for the $\nu_{\mu} \rightarrow \nu_{e}$ transition probability ($P_{\mu e}$), with terms up to the second order in $\theta_{13}$, is given by the following equation.

$$
P_{\mu e} \simeq \underbrace{\sin^2 \theta_{23} \sin^2 2\theta_{13} \frac{sin^2[(1-\hat{A})\Delta]}{(1-\hat{A})^2}}_{C_0} + \underbrace{\alpha^2 \cos^2\theta_{23}\sin^22\theta_{12} \frac{sin^2(\hat{A}\Delta)}{\hat{A^2}}}_{C1} \\
$$

\begin{equation}\label{eq:P_mu_e_in_matter}
 + \underbrace{\alpha \sin2\theta_{13} \cos\theta_{13} \sin2\theta_{12} \sin(\Delta) \frac{\sin(\hat{A}\Delta)}{\hat{A}}\frac{sin[(1-\hat{A})\Delta]}{(1-\hat{A})}}_{C_-} \sin\delta_{CP}
\end{equation}

$$
 +\underbrace{\alpha \sin2\theta_{13} \cos\theta_{13} \sin2\theta_{12} \cos(\Delta) \frac{\sin(\hat{A}\Delta)}{\hat{A}}\frac{sin[(1-\hat{A})\Delta]}{(1-\hat{A})}}_{C_+} \cos\delta_{CP}, $$
where
$$\Delta m^2_{ij} \equiv m^2_{i} - m^2_{j},\;   \alpha\equiv\Delta m^2_{21}/\Delta m^2_{31},\;  \Delta \equiv \frac{\Delta m^2_{31} L}{4E}, $$

$$ \hat{A} \equiv \frac{A}{\Delta m^2_{31}},\:  A= \pm 2\sqrt{2} G_{F}N_{e}E . $$

The $\theta_{12}$, $\theta_{23}$, $\theta_{13}$ are the mixing angles between neutrino mass eigenstates; $\Delta m^2_{31}$, $\Delta m^2_{21}$ are the neutrino mass square differences and $\delta_{CP}$ is the complex phase - the only parameter which can be responsible for the CP symmetry violation in neutrino sector. 
The $G_{F}$ is the Fermi coupling constant, $N_e$ is the electron density along neutrino propagation (inside the Earth in the long baseline experiment case) and $E$ is the neutrino energy; the positive sign in $A$ is for neutrinos and the negative one for antineutrinos.

The leading $C_0$ term in equation \ref{eq:P_mu_e_in_matter} contains the $\theta_{13}$ mixing angle and the largest matter effect dependency, the $C_1$ term is dominated by the solar $\theta_{12}$ mixing angle. The second order term in equation \ref{eq:P_mu_e_in_matter} marked as $C_{-}$ contains the $\sin\delta_{CP}$ dependency and is CP sensitive, it is negative for neutrinos and positive for antineutrinos therefore, makes the measurement of the $\delta_{CP}$ parameter possible; the last $C_{+}$ term depends on the $cos\delta_{CP}$ and is CP conserving. 
Neutrino oscillation probability changes significantly when a neutrino passes through dense matter. This is caused by the weak interaction which couples the neutrinos with ordinary matter (i.e. electrons, protons and neutrons). This gives an additional contribution to electron neutrinos because of their charge current (CC) interactions with electrons, mediated by $W^{\pm}$ exchange. The additional matter-related potential is noted in the formula by $A$. This additional potential generates interesting experimental effects like neutrino-antineutrino asymmetry, sometimes called "fake CP asymmetry". The matter effects increase the probability of $\nu_{\mu}\rightarrow \nu_{e}$ transition in case of normal mass hierarchy and decreases for inverted hierarchy. These effects need to be very well understood which makes the $\delta_{CP}$ measurement even more challenging.

To perform the $\delta_{CP}$ measurement, the best oscillation channel to observe is the $\nu_{e}$ appearance ($\nu_{\mu} \rightarrow \nu_{e}$), as  it contains a neutrino-antineutrino (CP) sensitive term. The $\nu_{\mu}$ disappearance channel ($\nu_{\mu} \rightarrow \nu_{\mu}$) can not give any information about the CP conservation or violation, however it is very useful to reduce the systematic errors. The further study focuses on these two oscillation channels as the most interesting ones for future long baseline oscillation experiments.

\begin{figure}[h!tb]
\centerline{%
\includegraphics[width=11.5cm]{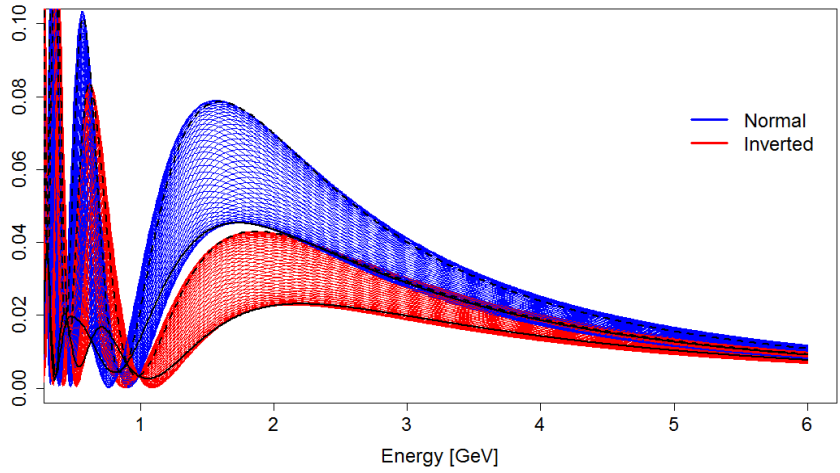}}
\caption{ The $\nu_e$ appearance probability $P_{\mu e}$ calculated as a function of the neutrino energy for a distance of 950 km and $\delta_{CP} \in(-\pi,\; \pi)$;  normal mass hierarchy - blue, inverted mass hierarchy - red; %$\delta_{CP}=\pi/2$ black solid line, \newline $\delta_{CP}= - \pi/2 $ black dashed line.
}
\label{Fig:Probability_wstega}
\end{figure}

For better illustration of the oscillation probability dependency on the value of the $\delta_{CP}$ parameter, Figure \ref{Fig:Probability_wstega} shows the calculated probability using the eq. (\ref{eq:P_mu_e_in_matter}) for the electron neutrino appearance after traveling a distance of 950 km through the Earth.  The calculations were performed for two mass hierarchies: normal and inverted, for $\delta_{CP}$ values from the full range i.e. $\delta_{CP} \in(-\pi,\; \pi) $. %Each thin line within a color band represents a different value of $\delta_{CP}$ with a $5$ [deg] step. 

For baseline 950 km (CERN-SUNLAB) the matter effects are big enough to slightly separate the normal (blue band) and inverted (red band) hierarchy scenarios in the neutrino energy range between the first and second oscillation maxima.

%%%%%%%%%%%%
%$\ldots$

Table \ref{tab:Global_fit_Schwetz}, \cite{Global_fit_3nu_oscillation} presents the values of the oscillation parameter obtained from the fit to the all available experimental data for the three flavour oscillation model. These values were used in the simulations performed for the experiment at SUNLAB.

\begin{table}[h!tb]
\centering
\begin{tabular}{|c|c|c|}
\hline
    & normal hierarchy & inverted hierarchy\\
   &  best fit point $\pm 1\sigma$ & best fit point $\pm 1\sigma$ \\
  \hline
  $sin^2 \theta_{12}$ & $0.304 ^{+0.013} _{-0.012}$ &  $0.304 ^{+0.013} _{-0.012} $\\
  \hline
  $sin^2 \theta_{23}$ & $0.452 ^{+0.052}_{-0.028}$ & $0.579 ^{+0.025}_{-0.037}$\\
    \hline
$sin^2 \theta_{13}$ & $0.0218^{+0.0010}_{-0.0010}$&  $0.0219^{+0.0011}_{-0.0010}$\\
  \hline
  $\frac{\Delta m^{2}_{21}}{10^{-5}\; eV^2}$& $7.50 ^{+0.19}_{-0.17}$ & $7.50 ^{+0.19}_{-0.17}$ \\
  \hline
   $\frac{\Delta m^{2}_{3l}}{10^{-3} \;eV^2}$ & $ +2.457^{+0.047}_{-0.047} $ & $ -2.449^{+0.048}_{-0.047}$\\
  \hline
  \hline
  $\delta_{CP} \;^{\circ} $ & $306 ^{+39}_{-70}  $ & $254  ^{+63}_{-62} $\\
  \hline
\end{tabular}
\caption{Values of the oscillation parameter obtained from the fit to the all available experimental data for the three flavour oscillation model. Second column contains values of parameters assuming the normal mass hierarchy ($\Delta m^2_{3l} \equiv \Delta m^2_{31} > 0 $) and the third one values of parameters assuming the inverted mass hierarchy  ($ \Delta m^2_{3l}\equiv \Delta m^2_{32} <0$). From \cite{Global_fit_3nu_oscillation}.}
\label{tab:Global_fit_Schwetz}
\end{table}

The software package GLOBES \cite{GLOBES} has been used in most of the simulation studies concerning neutrino oscillation experiments, in particular the accelerator long baseline ones. The whole simulation  process of neutrino oscillations for SUNLAB, from the source at CERN to the signal observed in the far detector, was described within GLoBES. GLoBES also provides the C-library and a number of functions to compute experimental features, such as expected event rates or systematic errors. However, in order to obtain the results of the oscillation experiment performance several external input data are needed. The most important ones are the neutrino flux and the detector response. In this study the neutrino flux was assumed as muon neutrino beam produced based on the CERN SPS accelerator and LAr-TPC detector response was estimated based on the ICARUS detector performance.

%%%%%%%%%%%%%%%%%%%%%%%%%%%%%%
\section{Neutrino beam}\label{sec:beam}

The definition of the neutrino source is an important first step on the way to simulate the neutrino oscillations in a long baseline experiment. In the case of the SUNLAB laboratory in Poland, like for other locations studied in Europe, the natural choice  of the origin of the neutrino beam was CERN. Therefore, in this study the neutrino beams based on  proton accelerators at CERN are considered. The distance from CERN to SUNLAB is 950 km, which defines the first oscillation maximum - the most interesting region to perform  the $\nu_{\mu}\rightarrow\nu_{e}$ appearance study - as corresponding to the neutrino energy of 1.92 GeV. For the $\delta_{CP}$ measurement, access to the second oscillation maximum  would also be interesting, so a wide-band beam  which covers neutrino energies down to about $ 400$ MeV should be considered.

The conventional procedure to obtain a neutrino beam is as follows: a
proton synchrotron delivers bunches of high-energy protons on a target, which
results in proton-target interactions and the production of secondary particles,
mainly pions and kaons. The intensity of the neutrino beam depends on the
number of delivered protons, so the commonly used unit to describe the neutrino
beam intensity is protons on target - p.o.t. Further, by using magnetic focusing
devices, such as a set of two cylindrically symmetric magnets called a horn and
a reflector, mesons with a selected charge sign and momentum are focused into a
decay tunnel. In the tunnel they decay mostly into a muon and muon neutrino.

The length of the decay tunnel should be chosen such that most of
the pions generating neutrinos will decay; however, the number of muon decays has to be 
kept low because among their decay products there are electron flavour neutrinos
which are the main background for  $\nu_{\mu} \; \leftrightarrow \; \nu_e$ oscillations.
The relatively rare three body decay of $K^+$ ($K_{e3}$ branching ratio ≈ 4.8\%)  is another
source of the $\nu_{e}$ background. At the end of the decay tunnel muons are usually monitored using dedicated muon detectors which indirectly provide information about the directions
of neutrinos.

The focusing system can be operated in the Positive Horn Focusing mode (PHF) and Negative Horn Focusing  mode (NHF). The produced beam is dominantly composed of $\nu_{\mu}$ ($\bar{\nu}_{\mu}$) if positively (negatively) charged mesons are chosen and focused. The background from other neutrino flavours consists of $\bar{\nu}_{\mu}$ (typically ~1-7\%), $\nu_e$ (usually $<1\%$), $\bar{\nu}_{e}$ (usually $< 0.1\%$) coming from the muon decays, $K_{e3}$ decays and also oppositely charged pions passing through the magnets. A good knowledge of the $\nu_{e}$ intrinsic beam contamination is crucial for
studies of $\nu_{\mu} \; \leftrightarrow \; \nu_e$ oscillations. At lower part of the beam energy spectrum the  $\nu_{e}$ contamination is mostly due to muon decays, while the kaon decays become more important at higher energies.

%\subsection{Neutrino beam simulation based on CERN-SPS}

 The simulation  of the neutrino beam used in this study is based on the Geant4 based software prepared by Andrea Longhin \cite{ALonghin} within the framework of the EUROnu and LAGUNA projects. As a result, the neutrino and antineutrino flux at a distance of 100 km are delivered in a file format suitable for the input to the GLoBES package. 
The main option for proton driver considered in this study is based on the existing and
very well understood SPS accelerator (Super Proton Synchrotron) delivering protons of energy 400 GeV under operation within the LHC acceleration complex. In the simulations the SPS proton beam power was assume to be 750 kW, resulting from the realisation of the LHC Injectors Upgrade project (LIU) \cite{LIU}.

\begin{figure}[h!tb]
\centerline{%
\includegraphics[width=6.4cm]{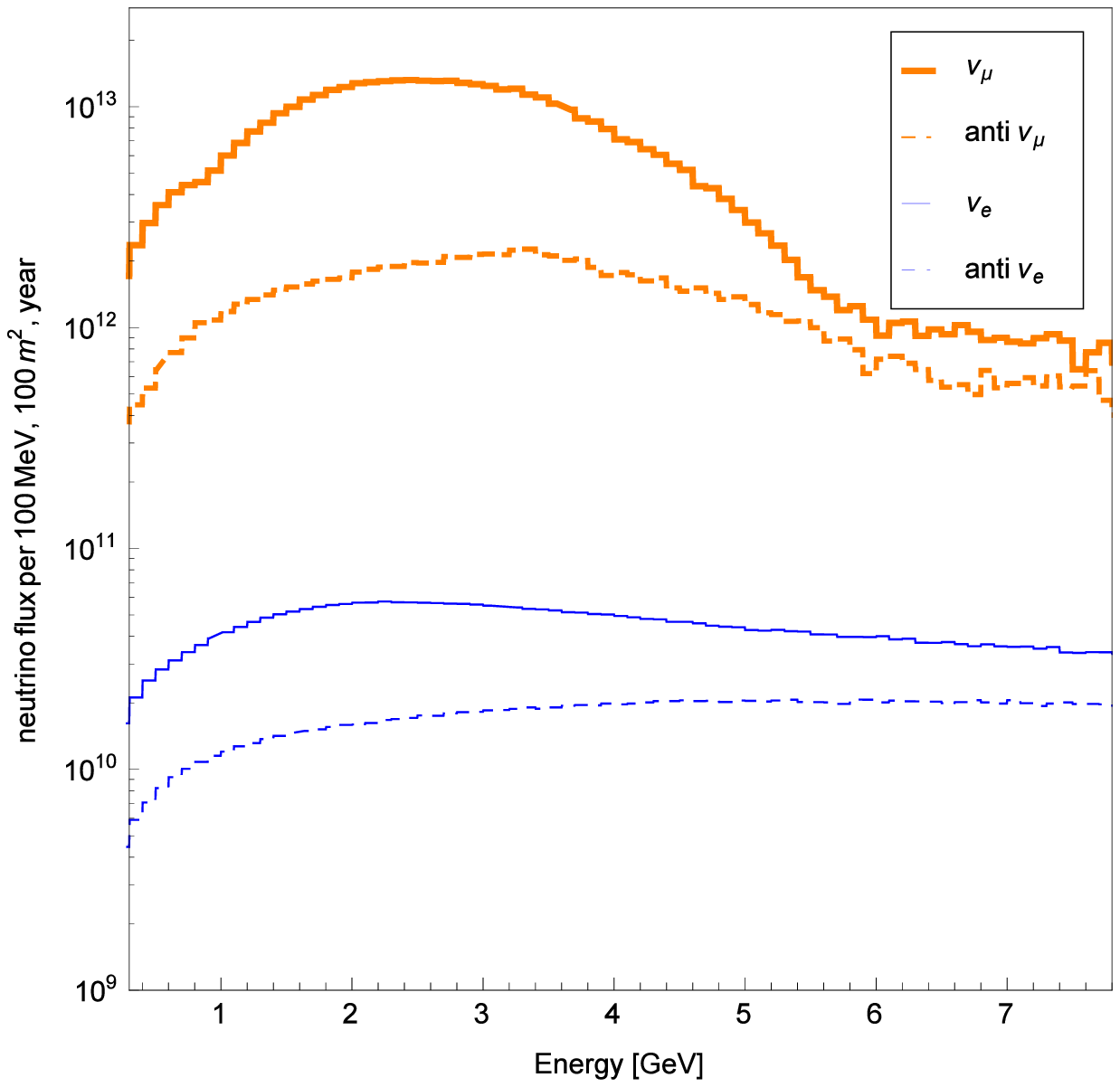}
\includegraphics[width=6.4cm]{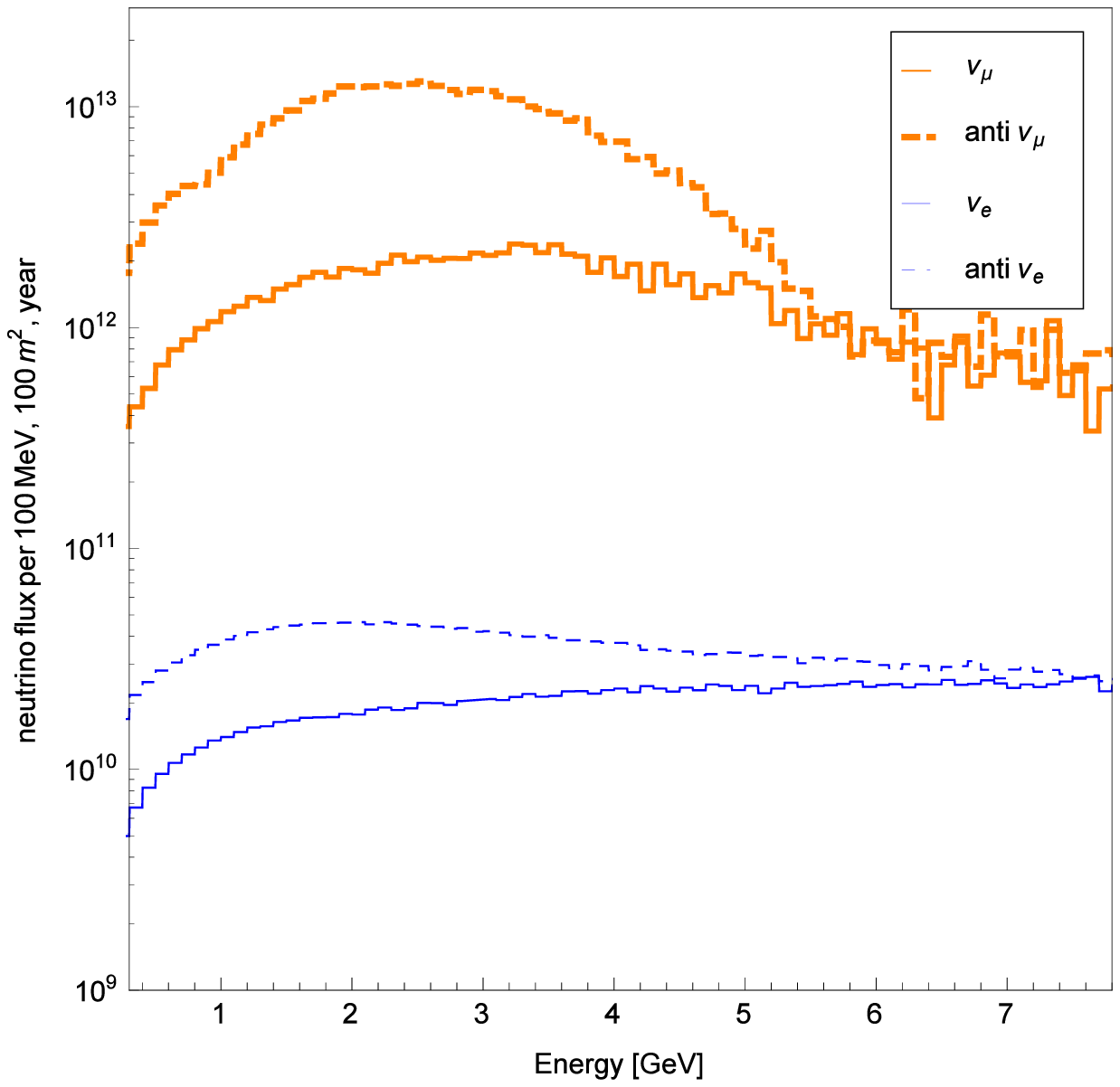}}
\caption{Fluxes in neutrino beam (left panel) and antineutrino enhanced beam (right panel) simulated for the
SUNLAB site using 400 GeV protons from SPS and assuming one year of running.}
\label{Fig:F2H}
\end{figure}

In addition, the beam had to be adjusted to the neutrino beam production, which is significantly different in the cycle structure from the beam prepared for the LHC. This corresponds to $ 7 \times 10^{13} $ protons extracted from the SPS every 6 seconds. The production of the wide band neutrino beam for the long baseline experiment presented in this paper is similar to the NuMI neutrino beam - Neutrinos at the Main Injector  at Fermilab. This kind of the design of the beam infrastructure is sometimes called in the literature 'NOvA-like'. This model has been chosen for this study,  because it is a relatively modern, well-tested setup. It consists of a thick graphite target, a focusing system of two thin magnetic horns named horn and reflector, and a hundred metres long meson decay tunnel. The target was modeled as a cylinder $ 1.0 $ m long, $ 4 $ mm in diameter of graphite with a density $ \rho = 1.85$ \textrm{$g/cm^{3}$}.
The horn focuses particles of one charge, e.g. those positively charged, and defocuses, i.e. rejects the particles of the opposite charge. In the horn particles with lower momenta are focused stronger, so the particles correctly charged, but with too low momenta may not reach the reflector and will be rejected from further focusing. The reflector bends the trajectories of particles which passed the horn and entered it to obtain a parallel narrow beam by giving additional focusing to high momentum particles and by protecting the low momentum particles from over-focusing. The mesons that pass the reflector are directed into a 100 m long decay tunnel where neutrino beam is produced. Figure \ref{Fig:F2H} presents the results of the beam simulation - the fluxes for neutrino (positive horn focusing mode) and antineutrino enhanced (negative horn focusing mode) beams. They include all neutrino flavours expected to appear in the beam. The flux normalised to the distance of 100 km after the decay tunnel is a direct input for GLoBES to perform further calculations.

%%%%%%%%%%%%%%%%%%%%%%%%%%%%%%%%%%%%%%%

%%%%%%%%%%%%%%%%%%%%%%%%%%%%%%%%%%%%%%
\section{Liquid Argon TPC}\label{sec:LArTPC}

The liquid Argon Time Projection Chamber (LAr TPC) is a relatively new and
very promising detector technique for neutrino physics \cite{Rubbia_TPC}. The LAr TPCs with
a mass up to 100 kton are studied in the context of the next generation long-baseline neutrino oscillation experiments, searches for proton decay and neutrino
astrophysics. Thanks to two kinds of signals from the ionization and scintillation
processes induced by charged particles in Argon, this technique offers a very fast trigger, excellent 3D event imaging and precise calorimetric measurements.
Due to these features even high energy neutrino events can be
registered and precisely measured. After the successful underground operation
of the ICARUS T600 detector \cite{ICARUS_2011} at the Gran Sasso Underground Laboratory, this detector technique has become the main option for the future long-baseline neutrino experiments. Currently some R\&D studies at Fermilab and CERN are carried on in view of their application in the LBNF-DUNE \cite{LBNF_DUNE} and SBN \cite{SBN} experimental programmes at Fermilab.
There are two types of LArTPC detectors developed for the neutrino experiments: single phase - ICARUS design and double phase - GLACIER design, and both are under further development. The difference between those two techniques is in the ionisation
signal treatment. In the single phase detector two or more planes of signal wires are immersed inside the liquid Argon volume to register the primary ionisation signal. This technique based on the ICARUS 600 ton detector, is being further developed under the WA104 programme \cite{WA104}, the operation of Proto-DUNE-SinglePhase detector \cite{ProtoDune_SP} at CERN and the SBN program at Fermilab in the LBNF-DUNE context. In the double phase detector primary ionisation electrons are extracted from the liquid to the gas phase, then multiplied by LEM (Large Electron Multiplier) devices and afterwards collected on the striped anode.  The double-phase-GLACIER design technique has not yet been demonstrated in a large scale and is being developed by the WA105 programme at CERN where the assembly of the ProtoDUNE-DualPhase detector is currently ongoing \cite{WA105_ProtoDune_DP},\cite{WA105}. 

%The work on the first large scale prototype
%(6m × 6m × 6m) is ongoing at CERN under the WA105 (LBNO-DEMO) programme with the %goal of its application to the DUNE detector as an alternative solution for one %or more detector modules.

%%\subsection{Single phase Liquid Argon TPC response - input for the simulation}

The far detector in the simulation of the SUNLAB-based oscillation experiment is assumed to be a
LAr-TPC single phase detector. Thanks to the ICARUS experiment, the information about the detector performance is known from the analysis of the experimental data and is not only based on simulations. Additionally, the Monte Carlo simulation program used by the ICARUS collaboration is well tuned to the detector response, which makes it a good tool for additional studies. However, one
should mention that the CNGS beam was optimised for the $\nu_{\tau}$ appearance study
with the mean neutrino energy of 17 GeV, so the ICARUS detector has collected higher energy data than are expected at the experiemnt at SUNLAB (first oscillation maximum at about 2 GeV). Therefore, some issues such as momentum reconstruction of not fully-contained long muon tracks will be a smaller problem for the  experiment at SUNLAB because of a much lower energy neutrino beam and a much bigger detector. The detector characteristics used as inputs in the sensitivity calculations with GLOBES are shown in Tables \ref{tab:LAr_detector_in_GLobes_nu_e_app} and \ref{tab:LAr_detector_in_GLobes_nu_mu_disapp}. Two detector masses assumed in the CP violation sensitivity simulations are 20 kton in the initial phase of the experiment and 100 kton in the final phase of the experiment. However, the detector mass is the scaling factor of exposure and in the first approximation it does not have a big impact on the event reconstruction quality. Therefore, the same tables are used for both mass values. In the analysis performed for SUNLAB two signal channels are included, the $\nu_e$ appearance and $\nu_{\mu}$ disappearance, so the parametrisation of signal and background treatment is given for both signal channels. For the $\nu_{e}$ appearance channel three types of background are taken into account as occurring in the LAr-TPC detector: the Neutral Current $NC$ events, muon neutrino Charge Current $\nu_{\mu} CC$ events wrongly identified as Charge Current electron events $\nu_{e} CC$ and finally $\nu_{e} CC$ events originating from an intrinsic beam contamination with $\nu_{e}$ neutrinos. For the $\nu_{\mu}$ disappearance channel the main source of background are the NC events, while the atmospheric $\nu_{\mu}$ neutrinos can be easily distinguished by the beam timing and direction.

\begin{table}[h]
\centering
\begin{tabular}{|l|c|}
  \hline
  Characteristics &  \\
  \hline
  signal  &  $ \nu_{e}$ appearance ($\nu_{e}\; CC$)\\
  background & $\nu_{\mu} \;CC$; $NC$; intrinsic $\nu_{e}\;(\nu_{e} CC)$\\
  \hline
  mass & 20 and 100 kton \\
  energy threshold & 300 MeV\\
 signal detection efficiency & 90\% for $e$ \\
  \hline
  \parbox{4.0cm}{energy resolution :\\ CC events } &  $0.15\sqrt{E/GeV}\;e^{\pm}$; $0.20\sqrt{E/GeV}\;\mu^{\pm}$,  \\
   NC events & Migration Matrix\\
  \hline
  \parbox{4.0cm}{background rejection: \\NC ($\nu_{\mu}\; NC$) }& 99.5\% \\
  from mis-id muons ($\nu_{\mu} \;CC$) & 99.5\% \\
  intrinsic $\nu_{e}$ & 20\%\\
  \hline
  norm. error for signal (syst.) & 5\%\\
  norm. error for background (syst.) & 5\%\\
  \hline
\end{tabular}\caption{Input for GLoBES for the $\nu_{e}$ appearance signal.}
\label{tab:LAr_detector_in_GLobes_nu_e_app}
\end{table}

\begin{table}[h]
\centering
\begin{tabular}{|l|c|}
  \hline
  Characteristics &   \\
  \hline
  signal  &  $\nu_{\mu}$ disappearance ($\nu_{\mu}\; CC$) \\
  background  & $NC$  \\
   \hline
  mass & 20 and 100 kton \\
  energy threshold & 300 MeV \\
 signal detection efficiency & 100\% for $\mu$\\
  \hline
   \parbox{4.0cm}{energy resolution:\\ CC events }&  $0.20\sqrt{E/GeV}$ for $\mu^{\pm}$\\
  NC events & Migration Matrix\\
  \hline
   \parbox{4.0cm}{background rejection:\\ $\nu_{\mu}\; NC$}& 99.5\% \\
   \hline
 norm. error for signal (syst.) & 5\% \\
norm. error for background (syst.)& 5\% \\
  \hline
\end{tabular}\caption{Input for GLoBES for the $\nu_{\mu}$ disappearance signal. }
\label{tab:LAr_detector_in_GLobes_nu_mu_disapp}
\end{table}
The additional systematic errors are due to the assumption that the total normalization of both the signal and background is allowed to vary by $5\%$ during the sensitivity calculations.

%%%%%%%%%%%%%%%%%%%%%%%
\section{Sensitivities of the oscillation experiment at SUNLAB for the $\delta_{CP}$ measurement}\label{sec:sensitivities}
The standard set of the oscillation parameters used in the $\delta_{CP}$ sensitivity calculations is:
$$
\theta_{12} = 0.601,\; \theta_{13}=0.162,\; \theta_{23} = 0.785,\\\;\;
 \Delta m^{2}_{21}= 0.0000762, \;\Delta m^{2}_{31} = 0.0024\;, $$

where values of mixing angles are given in radians and $\Delta m^2$ values in $\mathrm{eV^2}$. These are the central values of the oscillation parameters determined by the global fit \cite{Global_fit_3nu_oscillation}.  For $\theta_{23}$ the maximal mixing is assumed as the one still favoured by the atmospheric and T2K data. The Earth density is an important factor for incorporating matter effects in the simulations of the oscillation process. The Preliminary Reference Earth model (PREM) \cite{PREM}, which is a one-dimensional model providing the average description of the Earth properties as a function of its radius, has been applied. A single phase ICARUS type  LAr-TPC is assumed in two detector mass scenarios: 20 kton in the initial phase of the experiment and 100 kton in the final phase of the experiment. Ten years of running is assumed for each stage resulting in Exposure $= 20\;\mathrm{kt}\times 10\;\mathrm{yr} \times 10^{20} \mathrm{p.o.t.} = 2\times 10^{22}$ [$\mathrm{p.o.t. \times kt}$] for the initial stage and Exposure $= 10 \times 10^{22}$ [$\mathrm{p.o.t. \times kt}$] for the final phase. The neutrino mass hierarchy is assumed to be known and results are presented for both scenarios - normal and inverted hierarchies. The operation time is equally divided between the neutrino (PHF) and antineutrino (NHF) beam modes. The sensitivity for the CP-violation discovery potential is presented for both stages of the experiment.

%%\subsection{CP-violation discovery potential}

The potential to discover the CP-violation (CPV) for a
given value of $\delta_{CP}$, if it happens, is measured by the experiment ability to exclude the CP-conserving values, i.e. $\delta_{CP}= 0, \pm\pi$ at an assumed confidence level. By definition, this measurement becomes more difficult for the $\delta_{CP}$ values close to $0$ and $\pm\pi$. Therefore, the CP-violation may not be possible to discover if it is very small. The sensitivity for such a discovery is evaluated by minimizing the $\chi^2$ for a fit based on the predicted event rates for the assumed 'true' and 'test' values of $\delta_{CP}$ for the chosen oscillation channels. The 'true' parameters are chosen by Nature, the 'test' refers to the $\delta_{CP}$ values at which the likelihood is calculated with respect to the 'true' value. For the accelerator long baseline experiments the $\nu_{e}$ appearance and $\nu_{\mu}$ disappearance channels are considered and $\chi^2$ has the following form:

\begin{equation}\label{eq:Chi2}
    \chi^2 = \chi^2_{\nu_{e} app} + \chi^2_{\nu_{\mu} disapp}+ \chi^2_{syst}\;,
\end{equation}

where the $\chi^2_{\nu_{e} app}$  term corresponds to the $\nu_{e}$ appearance channel, the  $\chi^2_{\nu_{\mu} disapp}$ contains information from the $\nu_{\mu}$ disappearance channel, the $\chi^2_{syst}$ term  contains 5\% signal and background normalization errors.\newline

In order to evaluate the hypothesis that $\delta_{CP} = 0 \;or\;\pm \pi$, the $\Delta \chi^2$ was calculated in the following way:

\begin{equation}\label{eq:Delta Chi2}
    \Delta\chi^2 = \chi^2_{\delta_{CP}\;fix} - \chi^2_{\delta_{CP}\;vary} \;,
\end{equation}

where the $\chi^2_{\delta_{CP}\;fix}$ corresponds to the minimized $\chi^2$ value from Eq. (\ref{eq:Chi2}) at a fixed $\delta_{CP}$ value, while for $\chi^2_{\delta_{CP}\;vary}$  $\delta_{CP}$ varies over the full range of values ($\delta \in -\pi,\pi$).  Mass hierarchy is assumed to be known in the calculations,  either inverted or normal. The $\Delta \chi ^2$ is calculated with marginalisation over  all oscillation parameters except  $\delta_{CP}$, so the $\Delta \chi^2$ fit has one degree of freedom (1 dof). The CP violation hypothesis will be accepted at a given confidence level ($CL$) when the value of $\Delta\chi^2$ is below the chosen critical value. For example, for $\chi^2_{1 dof}$ the $CL =99.7\% $, equivalent to  $3\sigma$ for the Gaussian distribution, corresponds to $\Delta \chi^2 = 9 $, and $CL =99.99\%$,  equivalent to  $  5\sigma$, corresponds to   $\Delta \chi^2 = 25 $.

\begin{figure}[h!tb]
\centerline{%
\includegraphics[width=4.55cm]{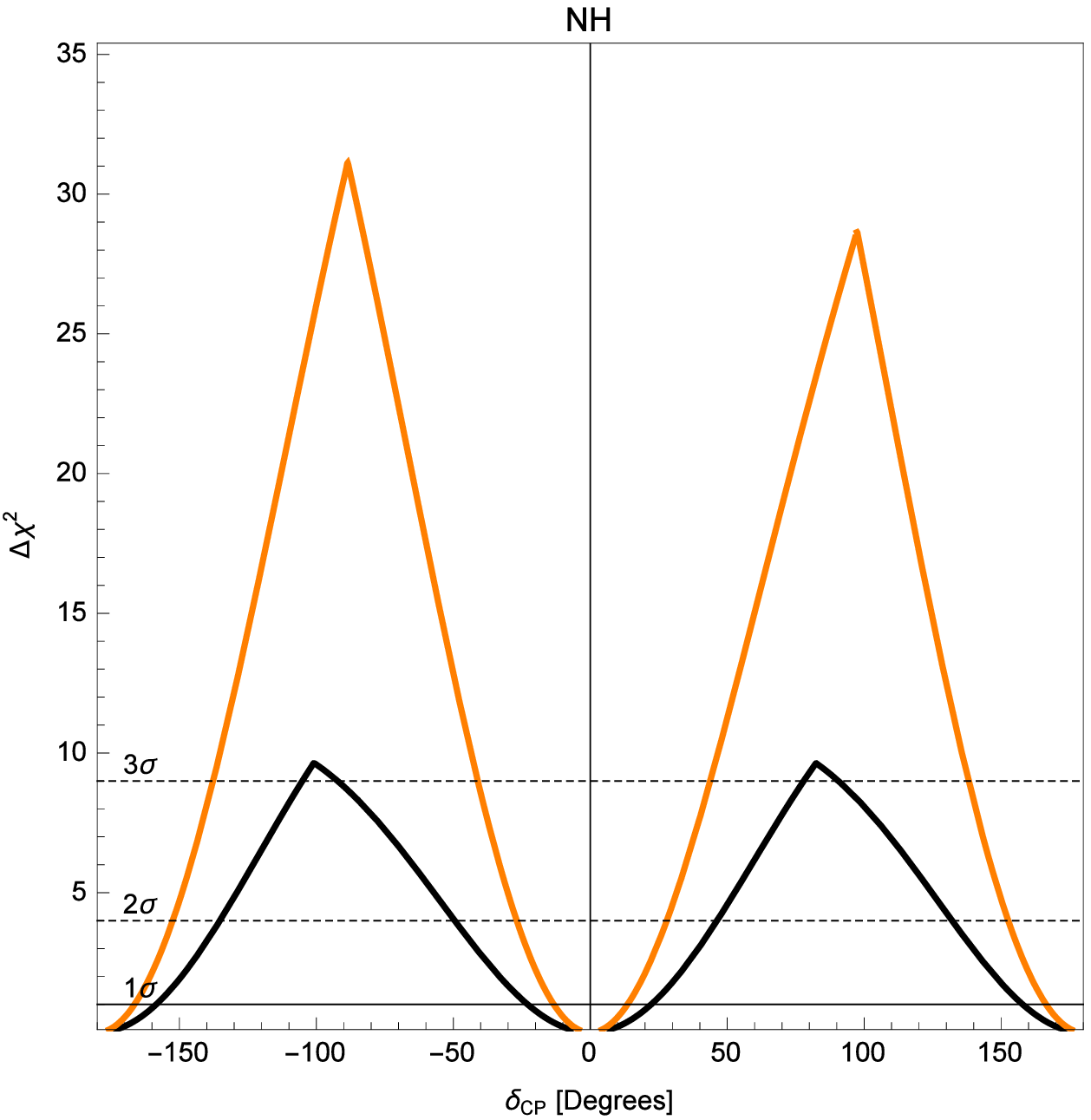}
\includegraphics[width=6.75cm]{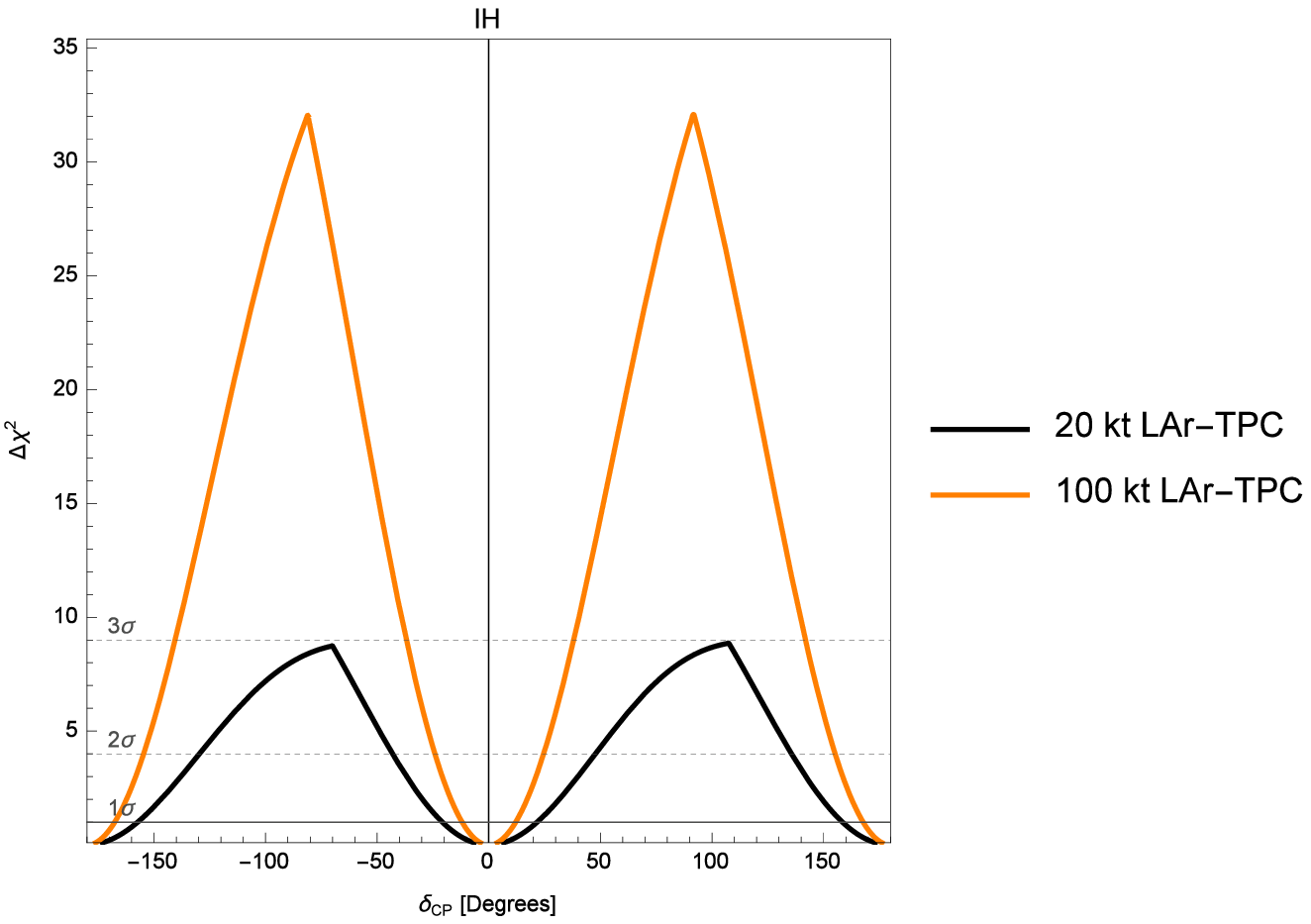}}
\caption{ Sensitivity for the CP violation discovery calculated for the SUNLAB site using neutrino beam originated from 400 GeV protons and assuming ten years of running, presented for assumed normal (left panel) and inverted (right panel) mass ordering for two LAr-TPC detector masses.}
\label{Fig:20vs100NH}
\end{figure}

The sensitivity plots for the CP violation discovery at SUNLAB, calculated using the GLoBES package, are presented in Figure \ref{Fig:20vs100NH}. The $\Delta \chi^2$ value calculated for the true $\delta_{CP} \in (-\pi, \pi)$ value and assuming the standard oscillation parameters is shown in each plot as a function of the true $\delta_{CP}$. The left panel of the figure is for normal hierarchy (NH) while the right panel is for inverted hierarchy (IH).  Horizontal lines depict the $\Delta \chi^2$ critical values for $1 \sigma$, $2 \sigma$, $3 \sigma$ of the CP violation discovery potential. During the study, two neutrino and antineutrino beam sharing scenarios: $50\%$ PHF + $50\%$ NHF and $40\% $ PHF + $60\%$  NHF were also tested; the latter was proposed to partially compensate for a smaller cross section for antineutrino interactions. However, the  $50\%$ PHF + $50\%$ NHF scenario gave a better sensitivity for the CPV determination and only this one is presented here.  Both the $\nu_{e}$ appearance and $\nu_{\mu}$ disappearance channels are included in the analysis. Even though the $\nu_{\mu}$ disappearance channel does not have a direct impact on the $\delta_{CP}$ measurement it  helps to improve the precision of the atmospheric parameters. The presented CP-violation discovery potential for the SUNLAB experiment in the initial and final phases can be summarised as follow. For the initial phase of the experiment, the CP-violation discovery potential at $2\sigma$ level covers 51\% of the $\delta_{CP}$ parameter range and reaches $3\sigma$ level in the case of NH for CPV close to maximal violation. The final phase of the experiment can give the coverage of $\delta_{CP}$ parameters of 58\% for NH and 60\% for IH at $3\sigma$ level and reaches $5\sigma$  in both cases for regions of $\delta_{CP}$ close to the maximal violation.

\section{Summary}\label{sec:summary}
The Polish site in the Polkowice - Sieroszowice mine, named SUNLAB, was one of the seven locations considered as future hosts for the large underground laboratory by FP7 LAGUNA project. Geomechanical feasibility studies for the SUNLAB laboratory, hosting a 100 kton liquid Argon GLACIER detector in anhydrite rock, were performed in the years 2008-2011 and gave very positive results. The neutrino oscillation experiment at SUNLAB was proposed with neutrino beam based on SPS proton driver, 950 km long baseline and 20 kton LAr-TPC as a far detector. This setup can achieve,  after 10 years of data taking the exposure of $ 2\times 10^{22}$ [$\mathrm{p.o.t. \times kt}$]. The study shows that for this design the CP violation discovery on the $5\sigma$ level is possible if the violation is close to maximal. The main limiting factor for the sensitivity of the experiment is the number of collected events. The better-suited proton drivers for neutrino beam production and large far detector allow achieving higher sensitivities for the CP violation discovery. The study prepared for the world leading long-baseline neutrino project DUNE-LBNF foresees sensitivities for the CPV discovery on similar level, only after 7 years of running. It assumes a new high-intensity beam from Main Injector at Fermilab and 40 kiloton LAr-TPC detector achieving the exposure of $ 9\times 10^{22}$ [$\mathrm{p.o.t. \times kt}$].

%%% The neutrino long baseline experiment coupled with the neutrino beam from CERN was the main motivation for LAGUNA and therefore the dedicated study for SUNLAB as a host of the far detector for the long baseline experiment was needed. 

%The LBNF project with the neutrino beam production based on Booster has become a flagship of the %research programme at Fermilab.%
%
%Now, the status of neutrino long baseline physics is more clear. It is known that a new large %underground laboratory in Europe will not be built in the near future, while the LBNF project in the %USA has become a flagship of the research programme at Fermilab.

\section*{Acknowledgments}
The study was partially funded by the NCN grant Preludium UMO-2011/03/N/ST2/01971. The simulations for this study were performed using Cracow Cloud One (CC1) \cite{CC1} infrastructure at IFJ PAN in Krak\'ow.

\vspace{2ex}%

\end{document}